\documentclass[twocolumn,showpacs,amssymb,aps]{revtex4}

\usepackage{graphicx}
\usepackage{psfig}
\usepackage{dcolumn}
\usepackage{bm}

\def\lessim{\lower.5ex\hbox{$\; \buildrel < \over \sim \;$}}
\newcommand{\pathnow}{}
\begin{document} \hbadness=10000
\topmargin -0.8cm\oddsidemargin = +0.1cm\evensidemargin = +0.1cm
\preprint{}

\title{{\rm PHYSICAL REVIEW C} 68, {\rm 061901(R) (2003)}\\[0.4cm]
Strange Pentaquark Hadrons in Statistical Hadronization}
\author{Jean Letessier}
\affiliation{Laboratoire de Physique Th\'eorique et Hautes Energies\\
Universit\'e Paris 7, 2 place Jussieu, F--75251 Cedex 05.
}

\author{Giorgio Torrieri, Steve Steinke, Johann Rafelski}
\affiliation{Department of Physics, University of Arizona, Tucson, Arizona, 85721, USA\\ }

\date{October 15, 2003, Published December 22, 2003}

\begin{abstract}
We study, within the statistical hadronization model, 
the influence of  narrow  strangeness carrying baryon resonances 
(pentaquarks) on  the 
understanding  of  particle production in relativistic
heavy ion collisions. There is a great
variation of expected yields as function of heavy ion collision energy due
to rapidly evolving chemical conditions at particle chemical freeze-out.
At relatively low collision energies, these new states lead to improvement of 
statistical hadronization fits. 
\end{abstract}

\maketitle
Enhanced production of strange
hadrons   in relativistic   
heavy ion collisions is well established~\cite{Man03,Zak03}. 
The availability  of a 
high abundance of strangeness favors production of
 strange hadron resonances, a topic of current intense
experimental interest in the field of relativistic heavy ion
 collisions~\cite{Raf01hp,Tor2001ue,Mar2002rw,Tor2002jp,Zha03dp,Mar03,Gau03}. 
The discovery   by the NA49 collaboration~\cite{Alt03}of a new $\Xi^{--}(1862)$ $I=3/2$ narrow $\Gamma<5$ MeV resonance 
in their $pp$ background, rather than in the $AA$ foreground data at
projectile  energy 158 GeV ($\sqrt{s_{\rm NN}}=17.2$ GeV) poses
the question in which conditions  one should look in heavy ion collisions 
for such new resonances.

This newly discovered  
hadron resonance has, given the mass and charge, an  exceedingly narrow width.
This  feature is common with  $\Theta^+(1540)$, another recently reported 
resonance~\cite{Nak03,Bar03,Step03,Barth03}, which
decays into the channel with quark content $uudd\bar s$ and  $I=0$. This is
believed to be  the predicted~\cite{Dia97}, lowest mass, pentaquark state~\cite{Sch03}.
The  $\Xi^*(1862)$ can be interpreted as its  most massive  isospin quartet member
$ssdd\bar u,\, ssud\bar u,\, ssud\bar d,\, ssuu\bar d$ 
with electrical  charge varying, respectively, from $-2$ to +1, in units of $|e|$.

Appearance of these new resonances can have many consequences in the field
of heavy ion collisions. 
We at first explore  how the introduction into the family of hadronic particles
of these two  new resonances,   $\Theta^+(1540)$ and $\Xi^*(1862)$,
influence the results of statistical hadronization fit to relativistic  heavy ion
hadron production experimental results. 
We use the same data set as has been employed in\ Ref~\cite{Zak03,Raf03ma,Raf03S}
and obtain predictions of how the relative abundances of these
new resonant states vary as function of the heavy ion collision energy. 

Importantly, only the two already identified states with $I=0$, and $I= 3/2$ 
of the anti decuplet, which also includes the  $I=1/2$, and $I= 1$ states 
are of relevance in the study
of the statistical hadronization fits. Thus, in our analysis, we do not depend 
on the unknown masses of  $I=1/2$, and $I=1$ states. However, 
 the interpretation of the newly discovered narrow  states as pentaquarks
enters our considerations decisively. The pentaquark valance quark content 
enters  the assigned chemical fugacities and 
phase space occupancies. The yield is proportional to the  
presumed spin degeneracy of the new states, 
taken to be two for  a spin 1/2 anti decuplet.

In our approach~\cite{Zak03,Raf03ma,Raf03S},
 as in other recent work~\cite{Bec03},  the chemical equilibrium 
and non-equilibrium is considered.
Accordingly, we allow  quark pair phase space 
occupancies, for  light quarks $\gamma_q\ne 1$, and/or 
strange quarks $\gamma_s\ne 1$~\cite{Let00b}. Since we 
study at SPS the  total particle multiplicities, and at RHIC 
the  central yields which can be considered  produced by 
rapidity-localized fireballs of matter, we  require in
our fits  balance in the
 strange and antistrange quark content~\cite{Let93hi}.
\begin{table*}[bt]
\caption{
\label{TqSPS} The chemical freeze-out  statistical parameters found for 
 nonequilibrium (left) and semi equilibrium (right) 
fits to SPS results. We show $\sqrt{s_{NN}}$,  
the  temperature $T$,  light quark fugacity $\lambda_q$, 
strange quark fugacity $\lambda_s$,
the quark occupancy parameters $\gamma_q$ and
$\gamma_s/\gamma_q$. Bottom line presents
the statistical significance of the fit. 
The star  (*) indicates for $\lambda_s$ that it is a value resulting from
strangeness conservation constraint. For $\gamma_q$ that there is 
an upper limit  to which the value converged,
$\gamma_q^2<e^{m_\pi/T}$ (on left), or that the value of $\gamma_q=1$
is set (on right).
}\vspace*{0.2cm}
\begin{tabular}{l|ccc|ccc}
\hline\hline
$\sqrt{s_{NN}}$\,[GeV]    &17.2   & 12.3  &8.75    & 17.2   & 12.3  &8.75 \\[0.1cm]
\hline
$T$\,[MeV]       &$135\pm3$    & $135\pm3$      &$133\pm2$       & $157\pm4$     &$156\pm4$         & $154\pm3$\\[0.1cm]
\hline
$\lambda_q$      &$1.69(5)$ &$1.98(6)$ &$2.56(6)$ &$1.74(5)$ &$2.03(7)$   &$2.69(8)$\\[0.1cm]
$\lambda_s$      &$1.23^*$      &$1.27^*$       &$1.31^*$      &$ 1.20^*$       &$1.24^*$        &$ 1.24^*$   \\[0.1cm]
\hline
$\gamma_q$          &$1.68^*$     &$1.68^*$        & $1.69^*$    & $1^*$    & $1^*$           & $1^*$ \\[0.1cm]
$\gamma_s/\gamma_q$  &$0.91(6)$ & $0.83(4)$  &$0.85(6)$  &$0.66(4)$    &$0.60(4)$  &$0.67(5)$ \\[0.1cm]
\hline
$\chi^2/$dof      &11.4/6  & 4.3/2      &2.3/4  & 23/7   &8.9/3    & 4.0/5  \\[0.1cm]
\hline\hline
\end{tabular}
 \end{table*}
\begin{table}[bt]
\caption{
\label{TqRHIC} The chemical freeze-out  statistical parameters found for 
 nonequilibrium (left) and semi equilibrium (right) 
fits to RHIC results. We show $\sqrt{s_{NN}}$,  
the  temperature $T$,  light quark fugacity $\lambda_q$, 
strange quark fugacity $\lambda_s$,
the quark occupancy parameters $\gamma_q$ and
$\gamma_s/\gamma_q$.  Bottom line presents
the statistical significance of the fit. 
The star  (*) indicates for $\lambda_s$ that it is a value resulting from
strangeness conservation constraint. For $\gamma_q$ that there is 
an upper limit  to which the value converged,
$\gamma_q^2<e^{m_\pi/T}$ (on left), or that the value of $\gamma_q=1$
is set (on right).
}\vspace*{0.2cm}
\begin{tabular}{l|cc|cc}
\hline\hline
$\sqrt{s_{NN}}$\,[GeV]    &200          & 130       &200        & 130    \\[0.1cm]
\hline
$T$\,[MeV]                &$142\pm5$    & $143\pm3$  &$159\pm6$    & $159\pm2$\\[0.1cm]
\hline
$\lambda_q$       &$1.051(9)$ &$1.069(8)$   &$1.052(9)$   &$1.067(8)$ \\[0.1cm]
$\lambda_s$      &$1.018^*$ &$1.023^*$   &$1.018^*$ &$ 1.023^*$ \\[0.1cm]
\hline
$\gamma_q$                &$1.62^*$&$1.63^*$& $1^*$ & $1^*$ \\[0.1cm]
$\gamma_s/\gamma_q$       &$1.23(12)$ & $1.32(5)$ &$1.013(6)$&$1.13(4)$ \\[0.1cm]
\hline
$\chi^2/$dof               &2.9/6        & 16.9/20    &4.6/7       & 32.7/21  \\[0.1cm]
\hline\hline
\end{tabular}
 \end{table}
 
There are two independent fit parameters when we assume complete chemical
equilibrium, the 
chemical freeze-out temperature $T$ and the light quark fugacity 
$\lambda_q=\sqrt{\lambda_u\lambda_d}=e^{\mu_b/(3T)}$. The baryochemical
 potential $\mu_b$ is the physical parameter controlling baryon density. 
Strangeness conservation fixes the strange quark fugacity $\lambda_s$ 
(equivalently, strangeness chemical potential, for more details 
 see, e.g., \cite{Zak03}).
Adding the possibility that the number of strange quark pairs is not
in chemical equilibrium, $\gamma_s\ne 1$, we  have  3 parameters, and
allowing also that light quark pair number is not in chemical 
equilibrium, we have 4 parameters. These three alternatives 
will be coded as open
triangles, open squares and filled squares, respectively,  in
all  results we present graphically.

\begin{figure}[t]
\vskip 0.3cm
\hspace*{-.2cm}\psfig{width=8.5cm,figure=\pathnow 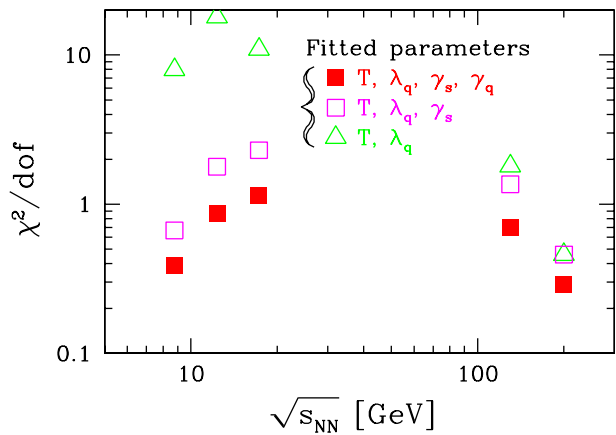}
\caption{\label{PLError} (Color online)
$\chi^2$/dof for statistical hadronization fits at SPS and RHIC: 
results are shown for 40, 80, 158$A$  GeV Pb on stationary Pb
target  collisions  and at 
RHIC for 65+65 and 100+100$A$ GeV Au--Au head on interactions. 
}
\end{figure}

 We find that  the  new resonance  $\Theta^+(1540)$  influences significantly the 
statistical hadronization fit to particle production at the lowest SPS 
energies. In a  baryon rich environment 
the introduction into the fit of  $\Theta^+(1540)$,  a 
$b=1$ baryon with `wrong' strangeness influences the strangeness 
balance condition, and thus indirectly the individual yields of all
strange hadrons.   
This leads to a   reduction in the statistical fit error 
for our hadronization study of the 40$A$ GeV Pb--Pb  reactions 
where we see a significant change in the relative yield of
kaons and $\Lambda$. We also find  changes in the details of the 
statistical fit parameters. In comparison to~\cite{Raf03ma}, 
aside  of the introduction 
of the new resonances, we also have harmonized our hadron
decay table with those used by the Krak\'ow group~\cite{Bar03kr}.
The improvement of the particle yield fit is both,  a   
theoretical confirmation of the validity of the statistical hadronization
model of particle production, and its applicability at low SPS energies.

We show how the fit error evolves in figure \ref{PLError},
which is also presented in the bottom lines of 
tables  \ref{TqSPS} and \ref{TqRHIC} along with the number of 
data points and resulting degrees of freedom. Considering  the small
number of degrees of freedom at SPS, we need
 $\chi^2/$dof $<1$ to have good significance of the fit.
The  errors seen in figure 
\ref{PLError} are, for the chemical nonequilibrium case (filled squares), 
sufficiently  small to allow us to conclude that  
the introduction of  $\Theta^+(1540)$  assures that the  statistical
hadronization works well down to the lowest SPS energies. 
To compare with  earlier results on $\chi^2/$dof, obtained prior 
to the discovery of these new resonances,  see Ref.~\cite{Raf03ma}, figure 16.

 An interesting point, seen in figure \ref{PLError}, is that  
the chemical equilibrium fit $\gamma_s=1,\,\gamma_q= 1$ is rendered 
unacceptable  at all SPS energies in presence of 
the new resonances. The semi-equilibrium fit, which 
allows a varying strangeness saturation, but assumes light quark 
equilibrium is generally resulting in twice as large  $\chi^2$
compared to the full non-equilibrium approach.  In a study of 
$\chi^2$ profile as function of $\gamma_q$ we find a clear 
and strong minimum for $\gamma_q\to \gamma_q^{\rm max}\equiv e^{-m_\pi/(2T)}$.
Acquisition by the fit of this limiting value implies that there is 
no fitting error in the $\gamma_q$ presented below.

\begin{figure}[!bt]
\vskip -0.5cm
\hspace*{-.2cm}\psfig{width=7.cm,figure=\pathnow 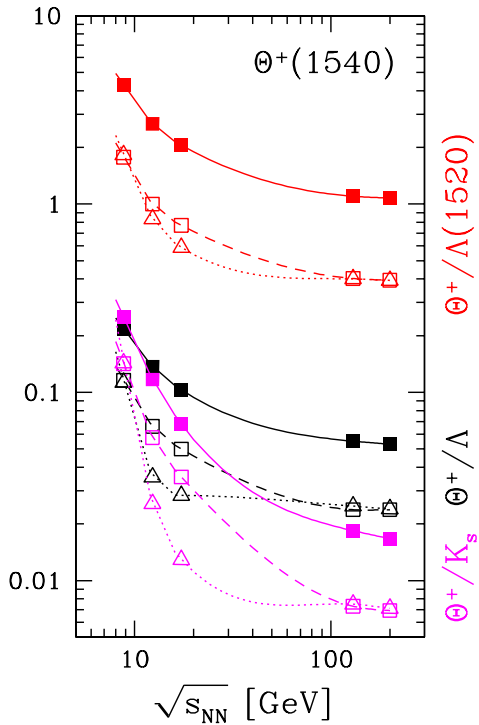}
\vspace*{-.2cm}
\caption{\label{PLQ1RATIO} (Color online)
Yield of $\Theta^+(1540)$ in relativistic heavy ion collisions, based on
statistical hadronization fit to hadronization parameters 
at SPS and RHIC  40, 80, 158$A$ GeV Pb on stationary Pb
target  collisions  and at 
RHIC for 65+65 and 100+100$A$ GeV Au--Au head on interactions. 
Relative yields with $K_s,\,\Lambda,$ and $\Lambda(1520)$ are shown
from bottom to top. 
}
\vskip -0.3cm
\end{figure}

The chemical freeze-out parameters of the fits considered play a
very important role in predicting the (relative) yield of hadronic particles,
and this dependence is even stronger for many pentaquark states, due to their
unusual quantum numbers.
These fit parameters for RHIC are shown in table \ref{TqRHIC}, and for 
SPS in table \ref{TqSPS} along with 
the freeze-out temperature. We note that for the full chemical
non-equilibrium, the freeze-out temperature is found to be
smaller than for semi-equilibrium case. This reduction is over-compensated
in pentaquark  yields by the significantly increased  value of $\gamma_q$.

\begin{figure}[!bt]
\vskip -0.5cm
\hspace*{-.2cm}\psfig{width=7.cm,figure=\pathnow 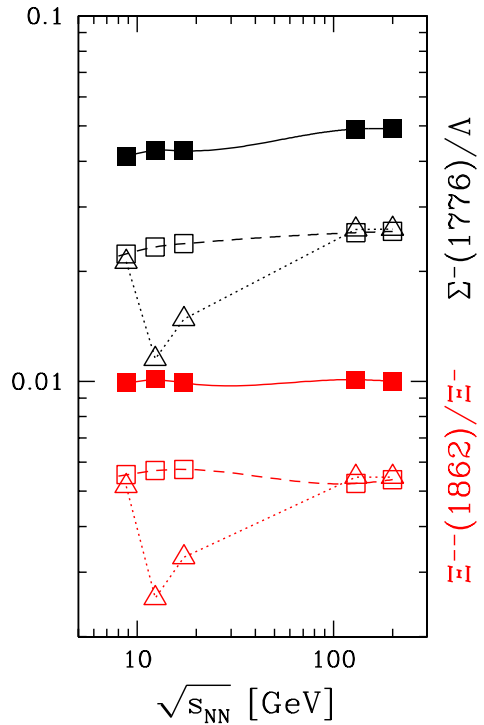}
\vspace*{-.2cm}
\caption{\label{PLQ34RATIO} (Color online)
Relative yields $\Xi^{--}(1862)/\Xi^-$ and $\Sigma^-(1786)/\Lambda$ are
shown from bottom to top, see figure \ref{PLQ1RATIO} for further details.
}
\end{figure}

We now consider the relative
yields of the new resonances in figures \ref{PLQ1RATIO} and \ref{PLQ34RATIO}. 
These yields vary strongly with collision energy 
for the case of  $\Theta^+(1540)$  in figure \ref{PLQ1RATIO}, but 
 are rather constant in figure \ref{PLQ34RATIO}. Certainly our result 
differs greatly from expectations arising from an earlier study of the 
statistical model production of the $\Theta^+(1540)$ resonance~\cite{Ran03fq}
where the  decisive variation of the particle yield with chemical potentials
was not explored. Moreover, the hadron yields, presented in \cite{Ran03fq},  did not
include the contributions from decay of short lived hadron resonances.
We checked that the relative particle yields shown  in~\cite{Ran03fq}
for zero chemical potentials and varying temperature 
 are mathematically correct, also as a cross check of our program. 

In figure \ref{PLQ1RATIO}, we show  (from top to bottom) the relative yields
 $\Theta^+(1540)/\Lambda(1520),\,\Theta^+(1540)/\Lambda,\,\Theta^+(1540)/K_s $
for chemical nonequilibrium (solid lines), semi-equilibrium
($\gamma_q=1$, dashed lines) and chemical equilibrium (dotted lines). 
The yields of $\Lambda$ used here include 50\% weak interaction cascade from $\Xi$.

The reason that the chemical nonequilibrium is  leading to 
greater than equilibrium yields is that the lower
hadronization temperature is overcompensated by the chemical
factors, e.g., 
$\Theta^+(1540)/\Lambda(1520) \simeq 1/2\, \gamma_q^2(\lambda_q/\lambda_s)^2$
ignoring the small mass difference. The factor $1/2$  is due to the difference in 
spin degeneracy. The actually observed  yield ratio $\Theta^+(1540)/\Lambda(1520)$
could be still greater since $\Lambda(1520)$ is seen at 50\% of the expected
statistical hadronization yield in heavy ion collisions. 
In figure \ref{PLQ1RATIO}, we also recognize that
the reason that there is such a significant impact
at low SPS energies of   $\Theta^+(1540)$ is that it
 is produced at the level
of +10--20\% of $\Lambda$ in  fits at 40$A$ GeV. 
This is due to the large prevailing baryochemical density. Clearly, this 
is the environment in which one would want to study the properties 
of this new resonance in more detail. However, at all energies considered, we
find that $\Theta^+(1540)$ is more abundant compared to $\Lambda(1520)$
and thus this new resonance could become an important probe of the hadronization
dynamics.

The observation of the  pattern of relative yield of 
$\Theta^+(1540)$,  seen in figure \ref{PLQ1RATIO},
would firmly confirm the 4-quark, one anti quark content of this state. Namely,
were for example the $\Theta^+(1540)$ another tri-quark baryon state, the
yield ratio with (strange) baryons would be quite flat as function
of collision energy. We further note that  the absolute
magnitude of the relative yield, seen in  figure \ref{PLQ1RATIO},
will be of help in establishing the degree of chemical equilibration.

In figure \ref{PLQ34RATIO}, we show at the bottom the expected relative 
yield of the $\Xi^{--}(1862)[ssdd\bar u]$ relative to $\Xi^-[ssd]$.
The 
$\Xi^*(1862)$ adds at the percentile level to the yield of observed $\Xi$
 and thus it  is less
influential in the statistical hadronization approach. 
The absence of variation of the relative yield with collision energy
is due to cancellation of chemical factors.
This relatively small relative yield 
at all collision energies here considered shows that indeed the $pp$ 
environment, where it has been identified by the NA49 collaboration, is 
most suitable. The dotted lines, in figure \ref{PLQ34RATIO},
are visibly breaking the trend in some of the results, indicating that
the large $\chi^2$ chemical equilibrium 
fit generates  unreliable statistical model parameters.

We also show, in figure \ref{PLQ34RATIO} on the top, 
the yield of the pentaquark state $\Sigma(1776)[sddu\bar u]$
which for purpose of this study is assumed at the mass indicated. 
Again due to cancellation of key chemical factors in  ratios shown 
in figure \ref{PLQ34RATIO}, 
both being proportional to $\gamma_q^2 \lambda_d/\lambda_u$, the
ratio is flat (except for the failed fit chemical equilibrium results).
Considering that  $\lambda_d\simeq \lambda_u$ and 
$\gamma_q\simeq 1.6$, the magnitude of relative yields 
seen in figure \ref{PLQ34RATIO}  is primarily due to the hadron
mass, and degeneracy. 

We have shown that inclusion of the  pentaquark states in the study of
particle production in heavy ion collisions improves the quality of our
fits to experimental data. We find that  
$\Theta^+(1540)$ state influences the low energy SPS particle yield fit results.
It can be expected that it will  be detectable, in particular 
 at  low heavy 
ion collision energies, and thus should become a new probe of 
hadronization dynamics. The other pentaquark states will be hard to 
observe in heavy ion collisions. 

\vspace*{.5cm}
Work supported in part by a grant from the U.S. Department of
Energy,  DE-FG03-95ER40937\,. LPTHE, Univ.\,Paris 6 et 7 is:
Unit\'e mixte de Recherche du CNRS, UMR7589.\\


\vskip 0.3cm

\end{document}